\documentclass{iaus}
\usepackage{graphicx}
\newcommand\Msun {M_{\odot}}
\title[Resolved Stellar Populations] 
{Resolved Stellar Populations\\ at the Distance of Virgo}

\author[Tolstoy]   
{Eline Tolstoy}

\affiliation{Kapteyn Astronomical Institute, University of Groningen,\\
Postbus 800, 9700AV Groningen, the Netherlands 
\break email: etolstoy@astro.rug.nl \\}

\pubyear{2005}
\volume{232}  
\pagerange{1--5}
\date{?? and in revised form ??}
\jname{The Scientific Requirements for Extremely Large Telescopes}
\editors{P. Whitelock, B. Leibundgut \& M. Dennefeld, eds.}

\begin{document}

\maketitle

\begin{abstract}
Top of the wish list of any astronomer who wants to understand galaxy
formation and evolution is to resolve the stellar populations of a
sample of giant elliptical galaxies: to take spectra of the stars and
make Colour-Magnitude Diagrams going down to the oldest main sequence
turn-offs. It is only by measuring the relative numbers of stars on
Main Sequence Turnoffs at ages ranging back to the time of the
earliest star formation in the Universe that we can obtain {\it
unambiguous} star formation histories. Understanding star formation
histories of individual galaxies underpins all our theories of galaxy
formation and evolution.  To date we only have detailed star formation
histories for the nearest-by objects in the Local Group, namely
galaxies within 700kpc of our own. This means predominantly small
diffuse dwarf galaxies in a {\it poor group} environment.  To sample
the full range of galaxy types and to consider galaxies in a high
density environment (where much mass in the Universe resides) we need
to be able to resolve stars at the distance of the Virgo
($\sim$17~Mpc) or Fornax ($\sim$18~Mpc) clusters.  This ambitious goal
requires an Extremely Large Telescope (ELT), with a diameter of
50$-$150m, operating in the optical/near-IR at its diffraction limit.
\end{abstract}

\section{Understanding Galaxy Formation}

To understand the formation of any galaxy we have to investigate their
stellar components, which carry a memory of the entire star forming
history of a galaxy.  We know that galaxies fall more or less into the
{\it Hubble Sequence}, but we don't really understand the details of
why and how massive galaxies find themselves separated into these
different classes. ellipticals are perhaps the biggest mystery because
we don't have an example that is easy to study in our backyard.  The
Milky Way is a prime example of a ``typical'' spiral or disc galaxy.
It consists of a flattened disc ($M_{gas} \sim 10^{10}\Msun$;
$M_{disc} \sim 6 \times 10^{10}\Msun$), with a central bulge
($M_{bulge} \sim 2 \times 10^{10}\Msun$) and a diffuse halo ($M_{halo}
\sim 3 \times 10^{9}\Msun$). We also have a second example of a spiral
galaxy in the Local Group, namely M31.  The stellar content of our
Milky Way and M31 appear markedly different. For example, 
the stellar population in the halo of M31 is more metal rich in the 
mean than our halo. However, globally
both these galaxies contain a large gaseous star-forming disc, a bulge
and a halo and both galaxies have significant populations of globular
clusters.

It is difficult to reach definitive conclusions on the general
formation path of massive galaxies with a sample of just two.
However, to enlarge our sample we need to move considerably further
away in distance, the Sculptor Group and the M81 group (both at about
2$-$3 Mpc distance) contain several more large spiral galaxies, but
still no elliptical galaxy.  For the nearest large elliptical galaxy
we have to go to NGC\,5128 (Centaurus~A, which is a very peculiar galaxy,
S0+ Spec) at 3.5~Mpc distance, but unfortunately as this system is
very complex it is unlikely to be representative. The Leo Group at
$\sim$10~Mpc distance contains the nearest nearly normal elliptical
galaxy, NGC\,3379 (E1 or S0).  But of course the Virgo cluster is the
real prize for studying elliptical galaxies.  Virgo at an average
distance of 17~Mpc, with over 2000 member galaxies of all
morphological types, is the nearest large cluster of galaxies. It has
a sequence of bright elliptical galaxies that was already cataloged
by Messier in the 18th century, and a luminosity sequence reaching
down the smallest dwarf elliptical (e.g., Gavazzi {\it et al.} 2005), such as
NGC\,4486B, believed to be an analog of the Local Group compact dwarf
elliptical, M32.  It is only in the Virgo cluster that we can truly
sample the population variations of elliptical galaxies.

Studying galaxies in Virgo also allows us to sample all types of
galaxy in a {\it dense} region of the Local Universe (e.g., Hudson
1993), where we believe that the majority of galaxies in the Universe form and
evolve. This will allow us to study the differences in the
evolutionary progress of galaxies in dense versus diffuse
environments.

If we can't observe resolved stars in Virgo we are neither sampling
the stellar population of elliptical galaxies nor the high density
regions of the Universe.

\section{Resolving Individual Stars in Elliptical Galaxies}

It has been shown many times in the past that it is not possible to
determine accurate star formation histories going back to the oldest
times without resolving stars on the oldest main sequence turnoffs
(M$_V = +2$ to $+4$). This is because the same age-metallicity
degeneracy which plagues accurate interpretation of integrated spectra
is equally a problem for the accurate interpretation of the age and
metallicity variation of a stellar population based {\it solely} upon
photometry of stars on its red giant branch (e.g. Tolstoy 1998;
Aparicio \& Gallart 2004).

At the present time it is not possible to detect individual resolved
stars on the red giant branch in galaxies at distances greater than
$\sim$10 Mpc (with HST), and the oldest main sequence turnoff stars
(M$_V = + 4$) can only be detected out to about 700kpc distance (also
using HST).  Hence all studies of large galaxies, such as ellipticals
in dense environments (such as large clusters) have been made using
the integrated colours and spectra. This only gives limits on the {\it
average} age of the stellar populations and cannot provide an
accurate, detailed star formation history. Of course it is the best
tool we have, and it may work quite effectively if carefully applied,
but we have no way of testing it at present. There are considerable
on-going efforts to glean information about elliptical galaxy
properties from integrated spectroscopy, their spectra typically
resemble that of a metal rich (solar or higher) K-giant star, and so
there are likely to have been very few, if any stars formed in the
last 1-2 Gyr. There are tight correlations between indicators of age
and metallicity that suggest a limited range of both (e.g.,
J$\o$rgensen 1997). But the age-metallicity degeneracy make firm
conclusions highly uncertain, and model dependent (e.g., Worthey
1993).

Despite considerable efforts elliptical galaxies remain largely
mysterious, although some progress has been made. Commonly accepted
wisdom used to be that elliptical galaxies are simple systems:
gas-free, disc-free, rotationally flattened ellipsoids containing only
very old stars. As people have looked more carefully most of these
assumptions turn out to be wrong or at best crude approximations.
Massive elliptical galaxies are not flattened by rotation, but are
anisotropic; they do have an interstellar medium, but it is a very hot
medium (T $> 10^6$K) that is difficult to detect except in X-rays.  A
significant fraction of elliptical galaxies contain peculiar kinematic
components (such as counter-rotating cores) which suggest a {\it
violent} process of formation through merging of two or more systems
over their history. Furthermore, all ellipticals (and spiral bulges)
seem to contain super-massive black holes amounting to $\sim$0.2\% of
their mass.  It has also been seen that on careful examination
elliptical galaxies frequently contain faint stellar discs, and low
mass ellipticals sometimes appear to contain intermediate age stars.
But to understand properly and quantify these general inferences from
presently available data we need to resolve individuals stars in
elliptical galaxies. As most elliptical galaxies are
{\it predominantly} old we need to observe individual
stars on main sequence turnoffs down to the oldest ages (i.e., M$_V
\sim +4$). 

In Virgo there is a large range of elliptical galaxies to chose from,
both in type (dwarf, giant or S0) and luminosity.  For example, M87 is
one of the brightest elliptical galaxies in Virgo, it lies right in
the centre of the gravitational potential and X-ray emission of the
cluster.  M59 is another luminous elliptical which is considerably
less luminous than M87, but still one of the brightest elliptical
galaxies in Virgo. M59 contains around 2000 globular clusters, easily
10 times more than the Milky Way. It has a central surface brightness,
I$_0$ =15.8 mag arcsec$^{-2}$ (fairly typical for the giant elliptical
galaxies in Virgo).  M59 is one of the largest elliptical galaxies with
a half-light radius, R$_e = 39^{''}$ (3.25kpc).  NGC\,4486B, a dwarf
elliptical, on the other hand has an R$_e = 3^{''}$ (0.25kpc), and
Virgo contains the complete range in between.  The surface brightness
at R$_e$ for the giant elliptical galaxies are typically I(e) $\sim 20
- 23$mag arcsec$^{-2}$, and for dwarf elliptical galaxies this is a
very large spread, with M32 type objects such as NGC\,4486B with I(e)
$\sim 16.8$mag arcsec$^{-2}$, but the more diffuse dwarfs have I(e) $>
23 - 24$mag arcsec$^{-2}$.  The effective radius, R$_e$, varies from
63$^{''}$ for M\,87 to $< 5^{''}$ for galaxies such as NGC\,4486B,
i.e. from the more than 5kpc to less than 500pc. Thus an optimum
field of view for an instrument studying these galaxies should ideally
be larger than $2 - 3^{''}$ but need not be larger than about 1$^{'}$.

These surface brightnesses and angular sizes provide the requirements
for both resolution and sensitivity to make accurate colour-magnitude
diagrams of resolved stellar populations in elliptical galaxies at the
distance of Virgo.  If we cannot observe resolved stars in Virgo we
will not obtain a detailed understanding of elliptical galaxies, and
we will only get an accurate detailed understanding of elliptical
galaxies with complete star formation histories coming from {\bf Main
Sequence Turn-off} photometry of individual stars.

\begin{figure}[!ht]
\includegraphics[scale=0.68]{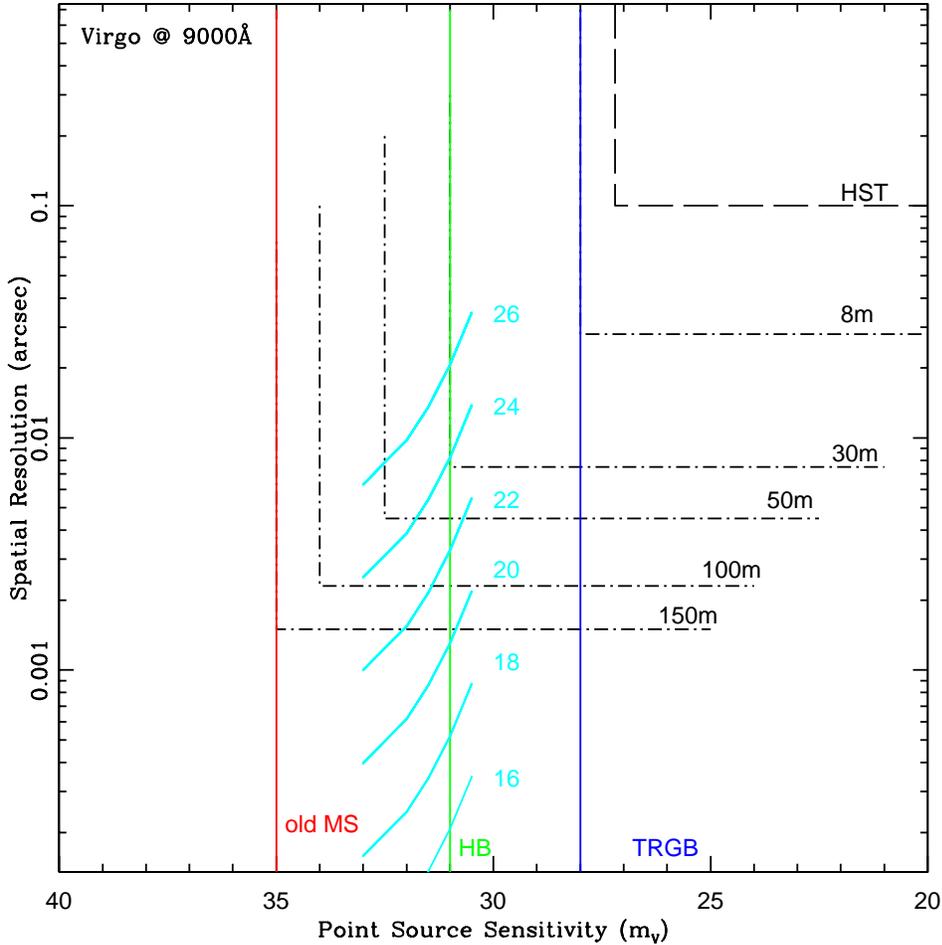}
  \caption{
The dashed half-box lines show the parameter space accessible to
different aperture telescopes at an observing wavelength of 9000\AA.
The spatial resolution is simply defined by the diffraction limit of
the telescope aperture. The point source sensitivity is determined
using the ESO-ELT exposure time calculator (etc.), and assuming a {\it
reasonable} exposure time (2$-$4 hours) and a moderate Strehl (30 \%).
As a `` reality check '' we calculated the same limits from the ELT
etc. for an 8m telescope.  We have also plotted the known limit of
HST/WFPC2 from experience observing resolved stellar populations in
the Local Group. The 3 vertical lines at m$_I = 35$, 31 \& 28
represent the approximate observed magnitude, at the distance of
Virgo, of the oldest, 13~Gyr, main sequence turnoffs (old MS); the
Horizontal Branch (HB) and the tip of the Red Giant Branch (TRGB),
respectively.  Also plotted are a series of sloping lines centres at 
m$_v \sim 31.5$ which represent the requirement to resolve the stars 
in an elliptical galaxy for the point source sensitivity at a variety 
of surface brightness levels. 
 }
\end{figure}

\section{Requirements: Limiting Magnitude \& Spatial Resolution}

As established in the previous section the requirements for making
accurate colour-magnitude diagrams at the distance of Virgo are
a combination of increasing spatial resolution with increasing 
sensitivity.

From deep HST observations of Leo\,A (Tolstoy {\it et al.} 1998), we can show
that with the 0.1\,arcsec pixels of the Wide Field Camera chip on WFPC2
we were {\it easily} able to carry out accurate photometry of
stars down the level of the Horizontal Branch (m$_V = 25$) for the
centre of this galaxy with a surface brightness of $\Sigma_V =
24.5$ mag arcsec$^{-2}$. 
This resulted in $\sim$2500 stars per 1$\times$1 arcmin WF chip
($200\times200pc^2$ at the distance of Leo A), or a stellar density 
of 1 star
per 1.44 arcsec$^{2}$.  This is by no means crowded.  At a distance
of Virgo this same surface area is $2.5\times2.5$ arcsec$^2$. 
Which provides us
with a minimum required resolution of $<$0.05~arcsec to carry out
the same observation of the same type of galaxy. Obviously elliptical
galaxies have typically a considerably higher surface brightness, and
hence stellar density, than faint dwarf irregular galaxies. Using for
example the surface brightness of an elliptical galaxy in Virgo (e.g.,
M59, I(e) $\sim$ 22 mag arcsec$^2$, and in the centre, I(0) $\sim$ 16
mag arcsec$^2$), we can determine the difficulties in resolving the
individual stars, even assuming that we have the sensitivity to detect
them. At R$_e$ we would expect an average stellar separation of
$\sim15mas$ between stars at the Horizontal Branch (or 5000 stars per
arcsec$^2$), and in the centre, $\sim1mas$ average stellar
separation (1~million stars per arcsec$^2$), at the Horizontal 
Branch.

In Fig.~1 we have attempted to visualize these limits in a plot of
point source sensitivity versus spatial resolution making an
attempt to correct for variations in stellar population. The above
numbers assume a constant star formation rate until the present day,
but for an Elliptical galaxy it is more realistic to assume that
star formation stopped at least 5~Gyr ago. 
The series of
sloping lines which represent the requirement to resolve the stars in
an elliptical galaxy for a given point-source sensitivity at a variety
of surface brightness levels are determined using theoretical stellar
evolution models combined to create an artificial stellar population,
which is only used in this context to determine the number of stars
expected at magnitudes brighter than the point-source sensitivity
limit. All the model lines assume the same star formation history (all
stars $>$ 5 Gyr old, and solar metallicity). As the point source
sensitivity increases a larger number of stars are detected which
decreases the spatial resolution requirement. The associated number is
the surface brightness ($\Sigma_V$)) of the ``galaxy'' (or region
thereof) being ``observed''.  It is clear that no telescope so far
imagined will be able to resolve stars at the level of the
Horizontal Branch, let alone old main sequence turnoffs, in the
central regions of Virgo elliptical galaxies. However, with a 50m
diffraction limited telescope, with some effort, or better still a
100m telescope we can hope to detect {\it and} resolve horizontal
branch stars in Virgo elliptical galaxies at their effective radius at
9000\AA. These are extremely preliminary results which rest on a large
number of highly uncertain parameters. It is clear this calculation
should be carried out carefully for each different telescope and
instrument design. The next logical step is to output the star
formation histories here used merely to determine the number of stars
in a given area to create an image with realistic point spread
functions.

\section{Spectroscopy}

Stars retain a sample of the interstellar medium out of which they
were formed in their outer atmosphere; a gaseous envelope which
surrounds the nuclear burning core. This envelope is believed to
remain unaltered through time for low mass stars, and thus a ``message
in a bottle'' is passed to us giving detailed information about the
interstellar medium out of which this star was formed.  Many of these
stars have very long lifetimes, and some of those we see today will
have formed (relatively) soon after the Big Bang.  The central core of the
star provides a background light source shining though this
envelope and the absorption lines we see tell us about the chemical
composition of this envelope. The more lines we can observe the more
elements we trace the better we understand the message the star is
providing for us.  Different chemical elements tell us how this
interstellar medium was enriched (if at all!)  by previous generations
of stars in the period before this star was formed. For example, the
$\alpha$-elements (O, Ca, Mg etc) can tell us how many and what mass
of supernovae type II explosions polluted the interstellar medium
before this star was made. Other elements can tell us about the likely
contribution of supernovae type I and even stellar winds coming from
AGB stars (e.g., McWilliam 1997).

For the most detailed abundance analyses (e.g., Shetrone {\it et al.} 2003)
high resolution spectroscopy is required (R$\sim$40\,000) with a
reasonably high signal-to-noise ($\sim 40$).  With an optimistic
outlook this may be possible for red giant branch stars at the
distance of M31, and maybe even Cen~A with an ELT type telescope. This
would allow us to understand the detailed differences between the
stellar populations of M31 and our Milky Way as well as other galaxies
in the Local Group (e.g., Venn {\it et al.} 2004).

However,if you are able to detect old main sequence stars in a
Colour-Magnitude Diagram ($>2$Gyr old) you are also able to determine
metallicities for individual red giant branch stars from particular
strong sets of lines (e.g., the Ca~II triplet at $\lambda\lambda = $
8498, 8542, 8662 \AA) observed at intermediate resolution (R$\ge$few
thousand), with moderate signal-to-noise ($\ge$10). The Ca~II triplet,
for example, has been well calibrated to high resolution metallicity
scale (e.g., Cole {\it et al.} 2004; Battaglia {\it et al.} 2006, in prep.), and accurately
measure the [Fe/H] of a star (not [Ca/Fe]!).  Although this indicator
(and others) don't (yet) provide much detailed information on any
element other than Fe they still provide a valuable insight into the
chemical evolution of the galaxy which can provide vital information
to correctly interpret a star formation history, especially if
deep main sequence photometry is not possible.

Ca~II triplet metallicities are only meaningful in a statistical
sense, because they are individually representative of [Fe/H] at a 
particular instant in the history of the galaxy that is always changing, 
and so large samples of stars should be observed per 
galaxy to build up an accurate picture of this variation over the 
entire star forming history of the galaxy. 
This requires multi-object spectroscopy over the
same field of view as the imaging discussed in the previous section.
Imaging is of course also required to make finding charts.  The tip of
the red giant branch is M$_I = -4$; which at Virgo is m(I)=27. This
might be feasible with a 100m telescope, but it will most likely be
overly challenging for smaller aperture sizes.

\begin{acknowledgments}
The author would like to gratefully acknowledge support from a
fellowship of the Royal Netherlands Academy of Arts and Sciences, and
travel support from the Leids Kerkhoven Bosscha Fonds. Thanks to Mike Irwin
for critical comments and corrections.
\end{acknowledgments}

\end{document}